\documentstyle[12pt]{article}

\def\Journal#1#2#3#4{{#1} {\bf #2}, #3 (#4)}

\def\APNY{{\em Ann. Phys.} (NY)}
\def\IJMPA{{\em Int. J. Mod. Phys.} A}

\def\NPB{{\em Nucl. Phys.} B}
\def\PLB{{\em Phys. Lett.}  B}

\def\RMP{\em Rev. Mod. Phys.}

\newcommand{\llabel}[1]{\label{#1}%
 }
 
\newcommand{\bib}[1]{\bibitem{#1}%
 } 

\newcommand{\3}{\ss}
\newcommand{\absatz}{\vspace{1.5ex}\noindent}

\newcommand{\be}{\begin{equation}}
\newcommand{\ee}{\end{equation}}
\newcommand{\bea}{\begin{eqnarray}}
\newcommand{\eea}{\end{eqnarray}}

\newcommand{\e}{{\rm e}}
\newcommand{\tr}{{\rm tr}}

\newcommand{\ph}{\mid \mbox{phys} \rangle}

\newcommand{\xv}{\vec{x}}
\newcommand{\xvp}{\vec{x}_{\perp}}
\newcommand{\ix}{(\xv)}
\newcommand{\ixp}{(\xv_{\perp})}
\newcommand{\dezweixp}{d^{\:\!2}\! x_{\perp}\;}
\newcommand{\dedreix}{d^{\:\!3}\! x\;}

\newcommand{\Ut}{\tilde{U}}
\newcommand{\Utd}{\tilde{U}^{\dagger}}
\newcommand{\Av}{\vec{A}}

\newcommand{\Piv}{\vec{\Pi}}

\newcommand{\de}{\;\partial}
\newcommand{\den}{\partial}
\newcommand{\dev}{\vec{\de}}

\newcommand{\denp}{\vec{\den}_{\perp}}

\newcommand{\Hp}{{\cal H}_{\rm phys}}

\def\ZZ{{\mathchoice {\hbox{$\sf\textstyle Z\kern-0.4em Z$}}
{\hbox{$\sf\textstyle Z\kern-0.4em Z$}}
{\hbox{$\sf\scriptstyle Z\kern-0.3em Z$}}
{\hbox{$\sf\scriptscriptstyle Z\kern-0.2em Z$}}}}

\begin{document}
 
\begin{titlepage}
\begin{flushright}
hep-ph/9608390\\
FAU-TP3-96/12 \\
August 1996
\\
\end{flushright}
\vspace*{1.5cm}
\begin{center}
\LARGE{\bf Defects in Modified Axial Gauge QCD$_{3+1}$}
 \end{center}
\vspace*{1.0cm}
\begin{center}

{\bf Harald W. Grie\3hammer\footnote{Talk presented at the ``Workshop on
 QCD'' at the American University of Paris,
June 3rd -- 8th, 1996, and the conference ``Quark Confinement and the Hadron 
Spectrum II'' at Villa Olmo, Como, Italy, June 26th -- 29th, 1996. Email: 
hgrie@theorie3.physik.uni-erlangen.de}}

\vspace*{0.2cm}

{\em Institut f\"ur Theoretische Physik III, Universit\"at 
Erlangen-N\"{u}rnberg,\\
Staudtstra\3e 7, 91058 Erlangen, Germany}
\vspace*{0.2cm}

\end{center}

\vspace*{2.0cm}


\begin{abstract}
Magnetically charged vortex defects are shown to arise in canonically quantised
modified axial gauge QCD$_{3+1}$ on a torus. This gauge -- being an Abelian
projection -- keeps  only the
eigenphases as dynamical variables of the Wilson line in $x_3$-direction. A
mismatch between the identification of large gauge transformations before and
after gauge fixing is indicated. 
\end{abstract}

\vskip 1.0cm
\end{titlepage}

\setcounter{page}{2}

\setcounter{footnote}{0}
\newpage


One of the main issues in non-Abelian gauge theories
is the presence of redundant variables.  Eliminating them by
``gauge fixing'', one hopes to identify the relevant degrees of freedom, the
non-perturbative part of which may solve the outstanding questions in the low
energy r\'egime of these theories. Monopole field configurations in the Abelian
projection gauges~\cite{MAP,KSW} seem to be a useful device to explain 
confinement by the dual Meissner effect~\cite{MAP,tHo2,Man}.  

In this context, a Hamiltonian formulation of QCD is especially useful 
since it allows one to bear in mind all
intuition and techniques of ordinary
quantum mechanics; formulating the theory in terms of unconstrained,
``physical'' variables is the easiest way to render gauge invariant results
in approximations. The choice of a compact base manifold softens the infrared
problem and allows the definition of zero modes. 

Here, the formulation by Lenz et al.~\cite{LNT} of Hamiltonian QCD in the
modified axial gauge~\cite{Yabuki} on a 
torus $T^3$ as spatial manifold is used, in which -- in 
contradistinction to the naive axial gauge $A_3=0$ -- the eigenphases of 
the Wilson line/Polyakov loop in $x_3$-direction are kept as
dynamical variables. This gauge belongs to the Abelian projection gauges, but
the magnetically charged configurations found do not have particle
character. On the contrary, their appearance indicates a failure of gauge
fixing. The outline of these results is the goal of the present paper; a
more rigorous treatment may be found in references~\cite{hgtrento,hgprom} and a
forthcoming publication.  

\absatz
As is well known, the Hamiltonian of pure QCD in the Weyl gauge $A_0=0$,
quantised by imposing the canonical commutation relations 
between fields $\Av$ and momenta $\Piv$, allows for time independent gauge 
transformations whose infinitesimal generator is Gau\3' law, $G^a\ix=
\dev\cdot\Piv^a\ix + gf^{abc}\Av^b\ix\cdot\Piv^c\ix$.
It cannot be derived as an equation 
of motion in the Hamiltonian formalism but has to be imposed on states,
defining the physical Hilbert space $\Hp$ as 
\be
   \llabel{gauss}
   \forall \ph \in \Hp\;:\;G^a\ix\ph =0\;\;\forall a,\;\xv.
\ee
On a torus $T^3$ with length of the edge $L$, one imposes periodic boundary
conditions for all fields and derivatives as well as for the gauge
 transformations~\footnote{Fermions will not affect the arguments given
here. Their sole trace is not to allow for twisted boundary 
conditions~\cite{tHo}.}
\be
   \llabel{pbc}
    \Av(\xv) = \Av(\xv+L\vec{e}_i)\;\;,\;\;
    \Piv(\xv) = \Piv(\xv+L\vec{e}_i)\;\;,\;\;
    V(\xv) = V(\xv+L\vec{e}_i)\;\;.
\ee 
The functional space $\cal H$ whose subspace $\Hp$ is consists hence of 
the space of periodic functionals. Eq.~(\ref{pbc}) is closely related to the
vanishing of all total colour charges in the box and to translation 
invariance~\cite{hgprom,Bala}.  

``Gauge fixing'' corresponds to a coordinate transformation in field space
\be
   \llabel{coordtrafo}
   \Av\ix=\Ut\ix[\Av'\ix +\frac{i}{g}\dev]\Utd\ix 
\ee
to a basis splitting explicitly in unconstrained ($\Av',\;\Piv'$) and 
constrained ($\Ut,\;\tilde{\vec{p}}\,$) variables and respective
conjugate momenta. Any operator $\cal O'$ commuting with Gau\3' law   
does not contain redundant degrees of freedom and hence can be written as
(possibly complicated) function of $\Av'$ and $\Piv'$ only. 
Operators for which $[{\tilde{\cal O}}, G^a\ix]\not=0$ must contain variables
$\Ut,\;\tilde{\vec{p}}$ constrained by Gau\3' law. 

It is also well known that not all gauge transformations are generated by
(\ref{gauss}) but that the unitary operator $\Omega[V]$ of arbitrary,
time-independent gauge transformations $V\ix$ 
leaves physical states invariant only up to a phase
into which -- besides the winding number $\nu(V)$ -- 
the vacuum-$\vartheta$-angle
 enters as a new, hidden parameter:
\be 
   \Omega[V]\ph= \e^{i\vartheta\nu(V)}\ph
\ee
The winding number detector is the integral over the Chern--Simons three-form, 
\be 
  \llabel{wa}
       W[A] := \frac{g^2}{16 \pi^2} \int \dedreix \;\epsilon^{ijk}
	     \tr[ F_{ij}A_k + \frac{2i}{3}g A_iA_jA_k]\;\;,
\ee
since it commutes with Gau\3' law and therefore is invariant under small 
gauge transformations, but changes under large ones 
\be
    \llabel{trafowa}
    \Omega[V] W[A] \Omega^{\dag}[V] = W[A] + \nu(V)
              +\frac{ig}{8\pi^2}
	    \int \dedreix \;\epsilon^{ijk} \de_i \tr[A_j(\de_kV^{\dag})V]
\ee
by the winding number of the mapping $V\ix:T^3\to SU(N)$, 
\be
   \llabel{defn}
    \nu(V) := \frac{1}{24\pi^2} \int \dedreix \;\epsilon^{ijk}
	    \tr[(V\de_iV^{\dag})(V\de_jV^{\dag})(V\de_kV^{\dag})]
            \;\in \ZZ\;\;.
\ee    
The surface term in (\ref{trafowa}) vanishes because of the periodicity
condition  (\ref{pbc}). Every (and hence every
physical) configuration has mirror configurations from which it differs by
large gauge transformations. 

Since (\ref{coordtrafo}) behaves as a gauge transformation, it acts on $W[A]$ 
by
\be
   W[A]\stackrel{\rm g-fix}{\longrightarrow}W[A']+\nu(\Ut)
                     +\frac{ig}{8\pi^2}\;\epsilon^{ijk}\int \dedreix
                      \de_i\;\tr[A'_j(\den_k\Ut^{\dagger}) \Ut]\;\;.
\ee
The operator $\nu(\Ut)$ and the surface term, which depends explicitly on the
physical field,  must vanish since $W[A]$ commutes with Gau\3' law and hence
contains only physical variables, 
\be
   \llabel{wafix}
   W[A]   \stackrel{\rm g-fix}{\longrightarrow} W[A']\;\;,
\ee
so that the freedom to perform large gauge transformations remains and the 
vacuum-$\vartheta$-angle is kept track of.

The modified axial gauge~\cite{LNT,Yabuki} has often been chosen since for it,
the splitting into $\Ut$ and $\Av'$ for every configuration $\Av$ can be 
given concretely, enabling the construction of the Hamiltonian in
terms of primed variables~\cite{LNT}. The ``zero mode'' fields $A_3'\ixp$ 
obey the modified axial ``gauge condition''
\be 
   \llabel{gfix}
    A_3'\ixp\mbox{ diagonal}\;\;,\;\;\de_3
    A_3'\ixp= 0\;\;,
\ee
and must remain relevant degrees of freedom since $A'_3\ixp$ are the phases of
the gauge invariant eigenvalues $\exp igLA_3'\ixp$ of the Wilson line in
$x_3$-direction. This is also expressed in the fact that the solution to 
(\ref{coordtrafo}) 
cannot be given for $A_3'\ix=0$ since then $\Ut$ would not be periodic in 
 $x_3$-direction and hence one would leave the space $\cal H$ of periodic
functionals. Allowing for a colour diagonal zero mode, one finds as
$x_3$-periodic solution to (\ref{coordtrafo}) and (\ref{gfix})
\be
   \llabel{solnut}
   \Ut\ix= \mbox{P}\e^{ig \int\limits_0^{x_3} dy_3\;A_3(\xvp,y_3)}
	 \Delta\ixp\; \e^{-igx_3 A_3'\ixp}\;\;,
\ee
where $\Delta\ixp$ diagonalises the Wilson line,
\be
   \llabel{diag}
   \Delta\ixp\; \e^{igL A_3'\ixp}\;\Delta^{\dagger}\ixp=\mbox{P}\!\exp ig
   \int\limits_0^L dx_3\;A_3\ix\;\;. 
\ee
The Hamiltonian in this representation has been constructed by Lenz et 
al.~\cite{LNT}

The residual gauge transformations can either be constructed by explicit
transformation of $\Omega[V]$ to the transformed physical Hilbert 
space~\cite{hgprom} or by an inspection~\cite{LNT} of the freedoms in the
 solution (\ref{solnut}) of equation (\ref{coordtrafo}) defining $A'_3\ixp$.
$\Omega'[V]\Av'\ix\Omega'^{\dagger}[V]$ again has to obey the ``gauge
condition'' (\ref{gfix}), so that the freedom to perform gauge transformations
which mix diagonal and off-diagonal components of the fields $\Av'$ is removed
and the residual gauge group in the space of the modified axial gauge variables
is $[U(1)]^{N-1}$, the group of the Abelian projection 
gauges~\cite{MAP,KSW}. The displacements of the colour neutral fields
\be
   \llabel{displacements}
   \Omega'[V]\Av'_p\ix\Omega'^{\dagger}[V] = 
   \Av'_p\ix+\frac{2\pi}{gL}\vec{n}_p
\ee 
affect only the zero modes and will be of importance in what follows.
The $pp$-entry of the matrices $\Av'$ is denoted by $\Av'_p$, and 
$(\vec{n})_{pq}\in \delta_{pq}\ZZ^3$ is a vector consisting of diagonal,
traceless matrices in colour space with integer entries, fixed in this form by
the boundary conditions (\ref{pbc}). 

A word of caution is in order here. It is clear that (\ref{solnut}) is a 
{\em local} solution to (\ref{coordtrafo}),
i.e.\ that for a given $\xvp$, one can expect the eigenvalues of the
Wilson line to be the only physical variables. Still, that $\Ut\ix$ is also a
solution {\em globally}, i.e.\ that it can be chosen regular and 
periodic in all directions for {\em all} possible configurations and 
simultaneously for all points on $T^3$ is not self-understood. This is the 
crucial point in the following. 

\absatz
As seen above, if the modified axial gauge choice resolves only Gau\3' law --
namely  if (\ref{wafix}) holds -- one can start from the winding number 
operator $W[A']$ in order to identify all large gauge transformations for all 
(physical) configurations. With (\ref{trafowa}/\ref{displacements}) one obtains
\be
   W[A']\stackrel{\Omega'[V]}{\longrightarrow}
   W[A'] + \frac{1}{2}\sum\limits_p n_{i,p}\left[\frac{g}{2\pi L}\;
   \varepsilon^{ijk}\int \dedreix \den_j A'_{k,p}\ix\right]\;\;.
\ee
The surface term in brackets defines the total magnetic Abelian fluxes
$\Phi_i$ in the $i$-th direction through the box 
and is nonzero only if the diagonal gauge fields are non-periodic. This
suggests that the allegedly unconstrained variables of the modified axial gauge
are not operators on the Hilbert space of periodic functionals. But
they must be if $\Av'$ acts solely inside $\Hp\subset{\cal H}$. 
On the other hand, if $\Av'$ is sometimes (or even always) periodic, 
as (\ref{pbc}) may suggest, {\em no} large gauge transformations would exist 
in the allegedly physical Hilbert 
space of the modified axial gauge representation for certain configurations,
 in contradistinction to the above argument.

It turns out that the mismatch is due to the presence of magnetically charged
defects~\cite{hgprom}. Note from (\ref{coordtrafo}) and the periodicity of 
$\Av$ that $\Av'$ is periodic only if $\Ut$ is. Although the latter is by
construction periodic in the $x_3$-direction, and its leftmost term in 
(\ref{solnut}) is periodic in the transverse directions $\xvp$, the two other
terms are not~\cite{LNT,hgprom}. Firstly, the definition of the
phases $A'_{3,p}\ixp$ necessary to define the
rightmost term in (\ref{solnut}) involves a logarithm so that the eigenphases
might lie on different Riemann sheets at opposite boundaries,
\be   
   A'_3(\vec{x}+L\vec{e}_{i})-A'_3(\vec{x})= \frac{2\pi}{gL}
   \varepsilon^{ij}m_j\;\;\mbox{ for }  i=1, 2\;\;,
\ee
where $m_i$ is a diagonal, traceless matrix with integer
entries. The mapping $\exp igLA_3'\ixp\!:\,T^2\to \left[U(1)\right]^{N-1}$
splits into topologically distinct classes labelled by the winding 
numbers $m_i\in \ZZ^{N-1}$. On the other hand, assuming a lattice 
regularisation of the Jacobian of the coordinate transformation 
(\ref{coordtrafo})~\cite{LMT}, $A'_{3,p}$ is interpreted as 
azimuthal angle in $SU(N)$ and other Riemann sheets are inaccessible. 
Then, such defects will play no r\^ole for {\em dynamical} reasons.

Secondly, the diagonalisation matrix $\Delta$ (\ref{diag}) is
determined only up to right multiplication with a diagonal matrix.
Drawing from the technique of Gross et
 al.~\cite{GPY}, one can show~\cite{hgprom} that the diagonalisation can
be chosen continuous on $T^3$, but in general not periodic
because the mapping $\Delta\ixp\,:\,T^2\to SU(N)/[U(1)]^{N-1}$ decomposes into 
topologically 
distinct classes which are labelled by a diagonal, traceless matrix with
winding numbers as entries,  
\be
   m_{3,p}:=\frac{i}{2\pi}\int\dezweixp \denp\times\left(\Delta^{\dagger}\ixp
            \denp\Delta\ixp\right)_p\in \ZZ\;\;.
\ee
Here, the Jacobian plays no r\^ole and $m_3\not=0$ even when $A'_3$ is
periodic. 
The occurrence of this defect is completely analogous to the introduction of a
point singularity on each sphere $S^2$ about a 't Hooft--Polyakov monopole
when one tries to diagonalise the Higgs field, i.e.~to transform it to the 
unitary gauge. 

Like this Dirac string, the {\em magnetic vortex defects} which yield nonzero 
$\Phi_i$ in the modified axial gauge are not
physical particles, having neither mass nor position, but are artifacts of the
gauge chosen. These objects can therefore not condense, and the 't
Hooft--Mandelstam mechanism for confinement~\cite{MAP,tHo2,Man} cannot be 
associated with them 
(in fact, Lenz et al.~\cite{LMT} showed that gluons are confined in the strong
coupling limit due to the Jacobian).
The magnetic fluxes $\Phi_i$ obey the Dirac quantisation condition and have no
relation to 't Hooft's magnetic twist configurations~\cite{tHo}: The defects 
occur even when fermions are included since one started with strictly periodic
boundary conditions (\ref{pbc}) in $\cal H$.

As a result, $W[A']$ changes under residual gauge transformations by the
{\em field-dependent} number $\tr[\vec{n}\cdot\vec{m}]/2$. Therefore, the
local gauge fixing condition (\ref{gfix}) and its global counterpart
(\ref{wafix}) do not match. The variables $\Av'$ are not operators in the
original Hilbert space of periodic functionals, showing an inconsistency of
this gauge choice. At this moment, a gauge choice  in which the Wilson line 
is not diagonalised is under investigation with the expectations that this 
concretely solvable
gauge allows for large gauge transformations after the elimination of redundant
degrees of freedom and that it may clarify the relevance of the above
defects for physical questions. Nonetheless, the modified axial gauge may serve
as a starting point for an effective theory in which the defects have a ``long
lifetime''. A special r\^ole will then be played by the diagonalisation defect
$m_3\not=0$ since it is -- in twofold contradistinction to 
$\vec{m}_{\perp}\not=0$ -- only present in two transversal dimensions even 
when one chooses another compact transverse manifold like $S^2$ instead of 
$T^2$. They share this property with the large gauge transformations which are
also special to 3+1 dimensions. 

On the other hand, since only sparse concrete information on the
non-perturbative sector of QCD exists, the presence of large gauge
transformations and the associated vacuum--$\vartheta$--angle should
be carefully kept accounted for.

\section*{Acknowledgments}
I am grateful for intense discussions with F. Lenz and A.C. Kalloniatis.

\end{document}